\documentstyle[prd,aps,eqsecnum,preprint,epsf,floats]{revtex}
\newif\iftightenlines\tightenlinesfalse
\tightenlines\tightenlinestrue
\def\eslt{E\llap/_T}
\def\to{\rightarrow}

\def\tst{\tilde t}
\def\ttau{\tilde \tau}

\def\tg{\tilde g}

\def\tw{\widetilde W}
\def\tz{\widetilde Z}

\begin{document}
\draft
\preprint{\vbox{\baselineskip=14pt%
   \rightline{FSU-HEP-980626}\break 
   \rightline{UH-511-908-98}
}}

\title{THE REACH OF \\CERN LEP2 AND FERMILAB TEVATRON UPGRADES\\ 
FOR HIGGS BOSONS IN SUPERSYMMETRIC MODELS}
\author{Howard Baer$^1$, B. W. Harris$^1$, and Xerxes Tata$^2$} 
\address{
$^1$Department of Physics, 
Florida State University, 
Tallahassee, FL 32306-4350 USA
}
\address{
$^2$Department of Physics and Astronomy,
University of Hawaii,
Honolulu, HI 96822 USA
}
\date{July 4, 1998}
\maketitle
\begin{abstract}

Luminosity upgrades of the Fermilab Tevatron $p\bar p$ collider have
been shown to 
allow experimental detection of a Standard Model (SM) Higgs boson up to
$m_{H_{SM}}\sim 120$ GeV via $WH_{SM} \to \ell\nu b\bar{b}$ events.  This limit
nearly saturates the parameter space for many models of weak scale
supersymmetry (SUSY) with a minimal particle content.  It is therefore 
interesting to examine the SUSY Higgs reach of future Tevatron experiments.
Contours are presented of Higgs boson reach for CERN LEP2 and Tevatron luminosity
upgrades for three models of weak scale SUSY: the Minimal Supersymmetric 
Standard Model (MSSM), the minimal Supergravity model (mSUGRA) and a
simple Gauge Mediated SUSY Breaking Model (GMSB).  
In each case we find a substantial gain in reach at the Tevatron with
integrated luminosity increasing from 10~fb$^{-1}$ to 25-30~fb$^{-1}$.  
With the larger integrated luminosity, a Higgs search at the 
Tevatron should be able to
probe essentially the entire parameter space of these models. While a
discovery would be very exciting, a negative result would severely
constrain our ideas about how weak scale supersymmetry is realized.

\end{abstract}
\begin{center}
{\bf To appear in Physical Review D}
\end{center}


\section{Introduction}

One of the mysteries of elementary particle physics is the origin
of electroweak symmetry breaking (EWSB). In the Standard Model (SM),
EWSB occurs via the Higgs mechanism, a consequence of which is the
existence of a fundamental scalar particle, the Higgs boson $H_{SM}$
\cite{hhg}. Comparison of precision measurements of electroweak parameters 
with SM predictions indicates a preference for a light Higgs boson 
$m_{H_{SM}}=115^{+116}_{-66}$ GeV~\cite{ew}.

The Higgs boson has been searched for at collider experiments.  The 
current best limit on its mass is $m_{H_{SM}}>88$ GeV from searches for 
$e^+e^-\to ZH_{SM}$ at LEP2\cite{lep2}.  
The CERN LEP2 collider is expected to ultimately 
reach a center-of-mass energy $\sqrt{s}\simeq 200$ GeV and 
integrated luminosity of $100 \; pb^{-1}$ per experiment, allowing
an exploration of $m_{H_{SM}}$ up to approximately $107$ GeV.  
Experiments at the 
CERN LHC ought to be able to observe $H_{SM}$ for $m_{H_{SM}}< 800$~GeV,
although if $m_{H_{SM}}$ is in the intermediate mass regime
($m_{H_{SM}}\simeq 90-180$ GeV), several years of running may be required
to extract the $H_{SM}\to\gamma\gamma$ signal from QCD two photon
backgrounds.

It has been pointed out that a high luminosity Fermilab Tevatron
$p\bar p$ collider has significant reach for a Higgs boson\cite{stange}.
The most promising channel at the Tevatron is $q\bar{q}'\to WH_{SM}$,
where $W\to \ell\nu$ and $H_{SM}\to b\bar{b}$. Simulations of signal
and SM backgrounds\cite{tev2000,kuhlmann} (mainly $Wb\bar{b}$, 
$t\bar{t}$, $W^*\to t\bar{b}$, $tbq$ and $WZ$
production) have shown that a $5\sigma$ signal ought to be detectable 
above background if $m_{H_{SM}}=80-120$ GeV, provided that 
an integrated luminosity of 
25 fb$^{-1}$ can be accumulated at $\sqrt{s}=2$ TeV. 

In many particle physics models with weak scale supersymmetry (SUSY)
and the low energy particle content of the
Minimal Supersymmetric Standard Model (MSSM),
the lightest Higgs scalar $h$
has a mass that is typically $m_h\alt 120-125$ GeV\cite{kane}. 
Furthermore, frequently the lightest SUSY Higgs boson $h$ 
behaves much like the SM Higgs boson. 
Thus, the Higgs boson
mass reach of the Tevatron collider is
particularly fascinating in that it may nearly saturate the 
parameter space of many interesting supersymmetric models.
The implication is that, if SUSY exists, then high luminosity
upgrades of the Fermilab Tevatron 
$p\bar{p}$ collider will
either discover the lightest SUSY Higgs boson $h$, or will exclude
much of the parameter space of many SUSY models!

Our goal in this paper is to translate the already calculated Tevatron 
SM Higgs boson mass reach into a reach in parameter space of three
specific models involving weak scale supersymmetry. 
These models are used for most phenomenological analyses of supersymmetry.

The first model assumes the generic structure
of the Minimal Supersymmetric Standard Model (MSSM)\cite{mssm} with 
no assumptions about physics at scales beyond $\sim 1$ TeV.
In this case, we set all dimensional SUSY parameters (such as soft SUSY
breaking sfermion and gaugino masses, and $\mu$) to $m_{SUSY}=1$~TeV, 
except $A$-parameters which are set to zero, so that the relevant parameter 
space consists of
\begin{equation}
{\rm MSSM} :\ \{m_A,\ \tan\beta \}
\end{equation}
where $m_A$ is the mass of the pseudoscalar Higgs boson and
$\tan\beta$ is the ratio of Higgs field vacuum expectation values. 
Several papers have presented SUSY 
Higgs search projections for LEP2 and LHC in this parameter 
space\cite{baer,gunion,zwirner,barger,atlas,cms}. 

The second model we examine is the minimal 
supergravity (mSUGRA) model\cite{sugra}
with radiative electroweak symmetry breaking.  In this model, it is
assumed that SUSY breaking takes place in a hidden sector, and SUSY
breaking effects are communicated to the observable sector via
gravitational interactions. In the minimal rendition of this model, all
scalars have a common mass $m_0$ at the GUT scale, while all gauginos
have a common GUT scale mass $m_{1/2}$, and all trilinear scalar
couplings unify to $A_0$, where the universality of the various
parameters occurs at some ultra-high scale $\sim M_{GUT}$. Weak scale
sparticle and Higgs masses are obtained via renormalization group
running of soft SUSY breaking masses and couplings from $M_{GUT}$ down
to $M_{weak}$, where radiative electroweak symmetry breaking
occurs.  Ultimately, all sparticle masses and mixings are calculated in
terms of the parameter set
\begin{equation}
{\rm mSUGRA} :\ \{ m_0,\ m_{1/2},\ A_0,\ \tan\beta,\ sign(\mu ) \} 
\end{equation}
where $\mu$ is the superpotential Higgs mass parameter, whose magnitude is
fixed by the condition of radiative electroweak symmetry breaking.

The last model we consider is the simplest gauge mediated SUSY breaking
model~\cite{dine}. In this model, SUSY breaking again takes place in a
hidden sector, but SUSY breaking is communicated to the visible sector
via messenger fields which also interact via usual gauge interactions.
Sparticle masses are proportional to their gauge couplings, and their
overall scale is set by the parameter $\Lambda =\frac{F}{M}$, where
$\sqrt{F}$ is the SUSY breaking scale and $M$ the mass scale for the 
messenger particles. 
The model is parameterized in terms of~\cite{pedro,isajet}
\begin{equation}
{\rm GMSB} :\ \{ \Lambda,\ M_{mes},\ n_5,\ \tan\beta,\ sign(\mu ),\ 
C_{grav} \}
\end{equation}
where $n_5$ is the number of complete $SU(5)$ messenger multiplets ($n_5
\leq 4$ if $M_{mes}$ is $\leq 1000$~TeV),  and 
$C_{grav}$ is the ratio of hidden sector to messenger sector vacuum 
expectation values of auxiliary fields.

\section{Calculations}

These SUSY models are incorporated in the event generator 
ISAJET 7.37\cite{isajet}.  Therein the SUSY Higgs 
boson masses are calculated by minimizing the renormalization-group-improved 
one-loop effective potential. The minimization is performed at an optimized
scale choice $Q\simeq\sqrt{m_{\tst_L}m_{\tst_R}}$, which effectively includes
the dominant two-loop contributions\cite{hmass} to $m_h$.  We input SUSY 
parameter space values into ISAJET to calculate the various Higgs boson 
and SUSY particle masses and mixing angles,
as well as Higgs boson branching fractions to SM and SUSY 
particles\cite{bisset}.

\subsection{Calculations for CERN LEP2}

The associated and pair production cross sections of SUSY Higgs bosons at 
$e^+e^-$ colliders can be expressed as \cite{bisset,eehiggs1,eehiggs2}
\begin{eqnarray}
\sigma(e^+e^- \rightarrow Zh) & = & \sin^2(\alpha+\beta) \; \sigma_{SM} 
\nonumber \\
\sigma(e^+e^- \rightarrow ZH) & = & \cos^2(\alpha+\beta) \; \sigma_{SM} 
\nonumber \\
\sigma(e^+e^- \rightarrow Ah) & = & \cos^2(\alpha+\beta) \; \sigma_{SM} 
\; \overline{\lambda}_{Ah} \nonumber \\
\sigma(e^+e^- \rightarrow AH) & = & \sin^2(\alpha+\beta) \; \sigma_{SM} 
\; \overline{\lambda}_{AH}
\end{eqnarray}
where
\begin{equation}
\overline{\lambda}_{Aj} = \frac{\lambda^{3/2}(m_j^2,m_A^2,s)}
{\lambda^{1/2}(m_j^2,m_Z^2,s) [ \lambda(m_j^2,m_Z^2,s) +12m_Z^2/s ] }.
\end{equation}
They are written in terms of the SM result for associated Higgs 
production given by 
\begin{eqnarray}
\sigma_{SM} & \equiv & \sigma(e^+e^- \rightarrow ZH_{SM}) \nonumber \\
            & = & \frac{G_F^2 m_Z^4}{96 \pi s} (v_e^2+a_e^2) 
\lambda^{1/2}(m_j^2,m_Z^2,s) \frac{\lambda(m_j^2,m_Z^2,s) +12m_Z^2/s}
{(1-m_Z^2/s)^2}
\end{eqnarray}
with $\lambda(x,y,z)=(1-x/z-y/z)^2-4xy/z^2$, $v_e=-1+4\sin^2\theta_W$, 
and $a_e=-1$.

The initial state QED corrections to the above processes 
are sizeable \cite{eeqed1} and are therefore included in our analysis.  
This is accomplished by a resummation (exponentiation) 
of large logarithms due to soft photon emission\cite{eeqed2}.  
The final form is a convolution of the leading order cross 
section with a 
resummed piece:
\begin{equation}
\sigma(s)=\int_{M_H^2/s}^1 dx \; G(x) \; \sigma_{BORN}(xs)
\end{equation}
where the resummed piece 
\begin{equation}
G(x)=\beta (1-x)^{\beta-1} \delta^{S+V} + \delta^H(x).
\end{equation}
The soft/virtual and hard photon contributions represented by 
$\delta^{S+V}$ and $\delta^H$ may be be found in Ref. \cite{eeqed2} to 
${\cal O}(\alpha_{em}^2)$.  They are polynomials in $L=\ln s/m_e^2$. 
The term in the exponent $\beta=2 \alpha_{em} ( L - 1 )/ \pi$ 
where $\alpha_{em}=1/137.0359 \ldots$.

We obtain contours of LEP2 excluded regions for $e^+e^-\to Zh$ or $ZH$
searches by calculating the SM cross section for $ZH_{SM}$ production
for $m_{H_{SM}}=88$ GeV and $\sqrt{s}=183$ GeV\cite{lep2}, 
and then requiring that the corresponding $Zh$ and $ZH$ cross sections 
be less than this value.

The LEP2 ultimate reach contours are calculated by assuming LEP2 to run at 
$\sqrt{s}=200$ GeV, with each experiment accumulating $100$~pb$^{-1}$
of integrated luminosity. This has been projected~\cite{konst} to yield a 
discovery limit of about 107~GeV for $H_{SM}$ via the Bjorken process; 
we normalize our ultimate SUSY Higgs $Zh$ and $ZH$ cross sections
to the corresponding SM cross section. 

To obtain contours of the LEP2 excluded region from $e^+e^-\to Ah$
searches, we normalize the $Ah$ production cross section to the
exclusion contours presented in Ref. \cite{lep2ha}, which, for instance,
exclude $m_h=m_A=76$ GeV for $\tan\beta =20$ at $\sqrt{s}=184$ GeV.  We
also project the ultimate LEP2 reach via $hA$ production assuming that
discovery will be possible if the corresponding cross section at
$\sqrt{s}=200$~GeV is larger than
0.1~pb, while the charged Higgs boson will be detectable if it is
lighter than 95~GeV.

\subsection{Calculations for the Fermilab Tevatron}

Detailed simulations of the $p\bar{p}\to WH_{SM}+X\to\ell\nu b\bar{b}+X$
signal and background have been performed in Ref. \cite{tev2000} and
\cite{kuhlmann}, using the VECBOS, PYTHIA and HERWIG event generators.
The event selection cuts were as follows\cite{tev2000}:
\begin{itemize}
\item $p_T(\ell )>20$ GeV,
\item $\eslt >25$ GeV,
\item number of jets ($E_T>15$ GeV, $|\eta |<2.5$)$ \; \ge \; 2$,
\item number of jets ($E_T>30$ GeV, $|\eta |<2.5$)$ \; \le \; 2$,
\item $m_{min}\le m_{jj}\le m_{max}$,
\end{itemize}
where $m_{min}$ and $m_{max}$ are variable end-points for the dijet mass
bin (for $m_{H_{SM}}=100$ GeV, $m_{min}=84$ GeV and $m_{max}=117$ GeV, 
for example).  In addition, 
\begin{itemize}
\item the two jets are tagged via the CDF secondary vertex (SVX) 
$b$-tagging algorithm.
\end{itemize}
In Ref. \cite{kuhlmann}, the signal was increased by adopting the CDF 
standard soft lepton tag (SLT) and the CDF loose SVX $b$-tag 
to the second $b$-jet. Finally, some additional optimization of signal
to background was obtained by imposing the cut
\begin{itemize}
\item $|\cos\theta_H |<0.8$ ,
\end{itemize}
in the $WH_{SM}$ center-of-mass system. Tables for signal and background 
were presented for $m_{H_{SM}}=60$, 80, 100 and 120 GeV.

\begin{figure}
\centerline{\epsfbox{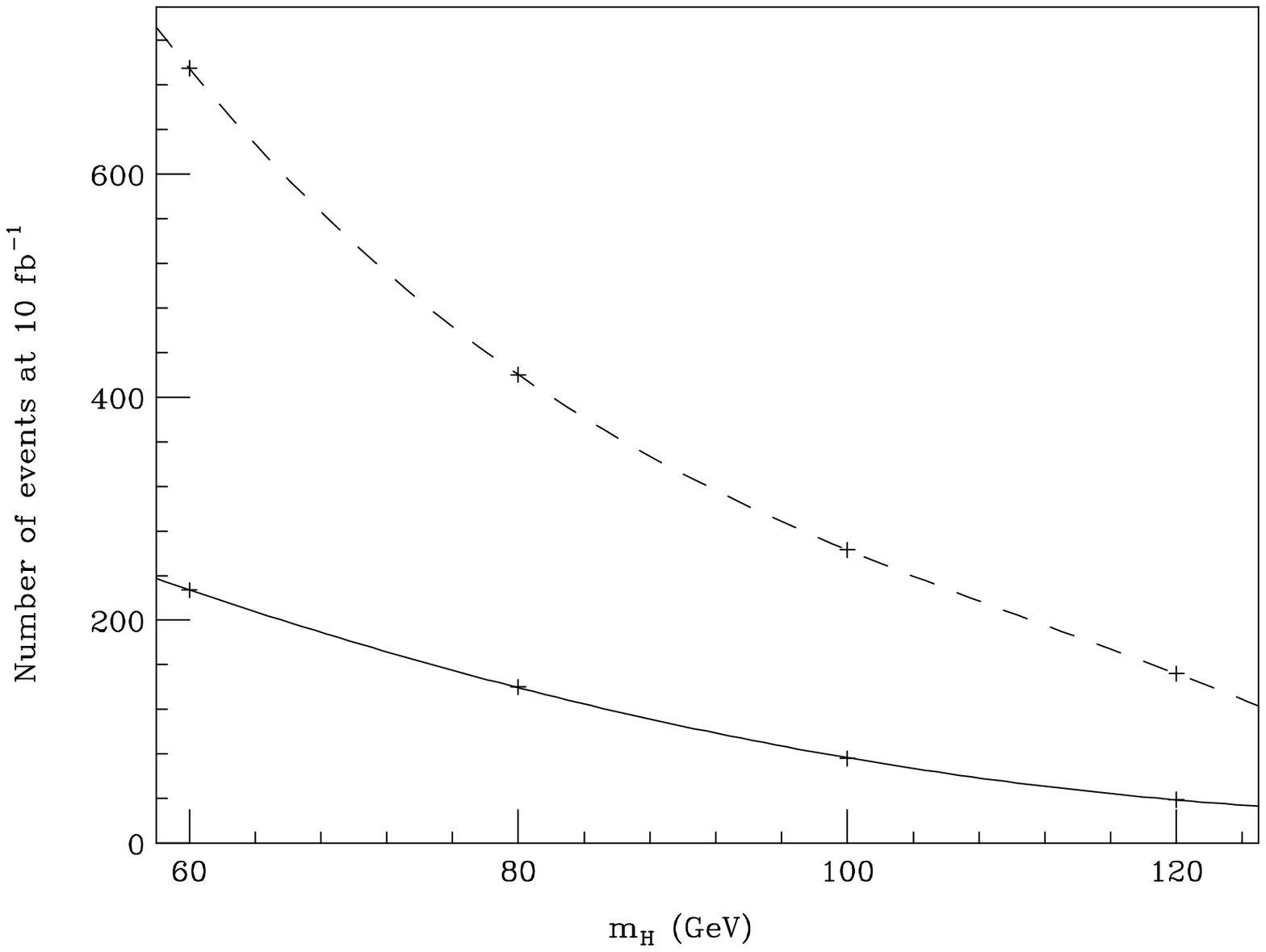}}
\vspace {1.0in}
\caption{}
\label{FIG0}
\end{figure}

We performed a simple fit to the signal presented in Table IV of 
Ref. \cite{kuhlmann}, and found that the number of signal events 
$S(m_{H_{SM}})$ in 10 fb$^{-1}$ could be represented by
\begin{equation}
\label{smsignal}
S(x)=640.6-8.765 x +0.03125 x^2 ,
\end{equation}
while the number of background events could be represented by
$B(m_{H_{SM}})$ with
\begin{equation}
B(x)=2948-62.6 x+0.5075 x^2 -0.0015 x^3 .
\end{equation}
The numerical results of Table IV of Ref. \cite{kuhlmann} and the above 
fitted curves are shown in Fig.\ \ref{FIG0}.

The supersymmetric $Wh$ and $WH$ production cross sections are given by
\begin{eqnarray}
\sigma(p\bar{p} \rightarrow Wh) & = & \sin^2(\alpha+\beta) \; \sigma_{SM} 
\nonumber \\
\sigma(p\bar{p} \rightarrow WH) & = & \cos^2(\alpha+\beta) \; \sigma_{SM} 
\end{eqnarray}
where $\sigma_{SM}=\sigma (p\bar{p} \to WH_{SM})$.  Thus, 
\begin{equation}
\sigma (p\bar{p} \rightarrow Wh \to \ell\nu b\bar{b})=\sin^2(\alpha+\beta) 
\frac{BF(h\to b\bar{b})}{BF(H_{SM}\to b\bar{b})} 
\left[ \frac{S(m_h)}{10 \; fb^{-1}} \right],
\end{equation}
with a similar equation for $WH$ production.  The SUSY and SM branching 
fractions are extracted from ISAJET and are given in \cite{bisset}.

As a criterion for observability at the Tevatron, we use $S/B>0.2$ and 
$S/\sqrt{B}>5$.  We note that $S/B \sim 0.3$ throughout most of parameter 
space.

\section{Results}

\subsection{Minimal Supersymmetric Standard Model (MSSM)}

\begin{figure}
\centerline{\epsfbox{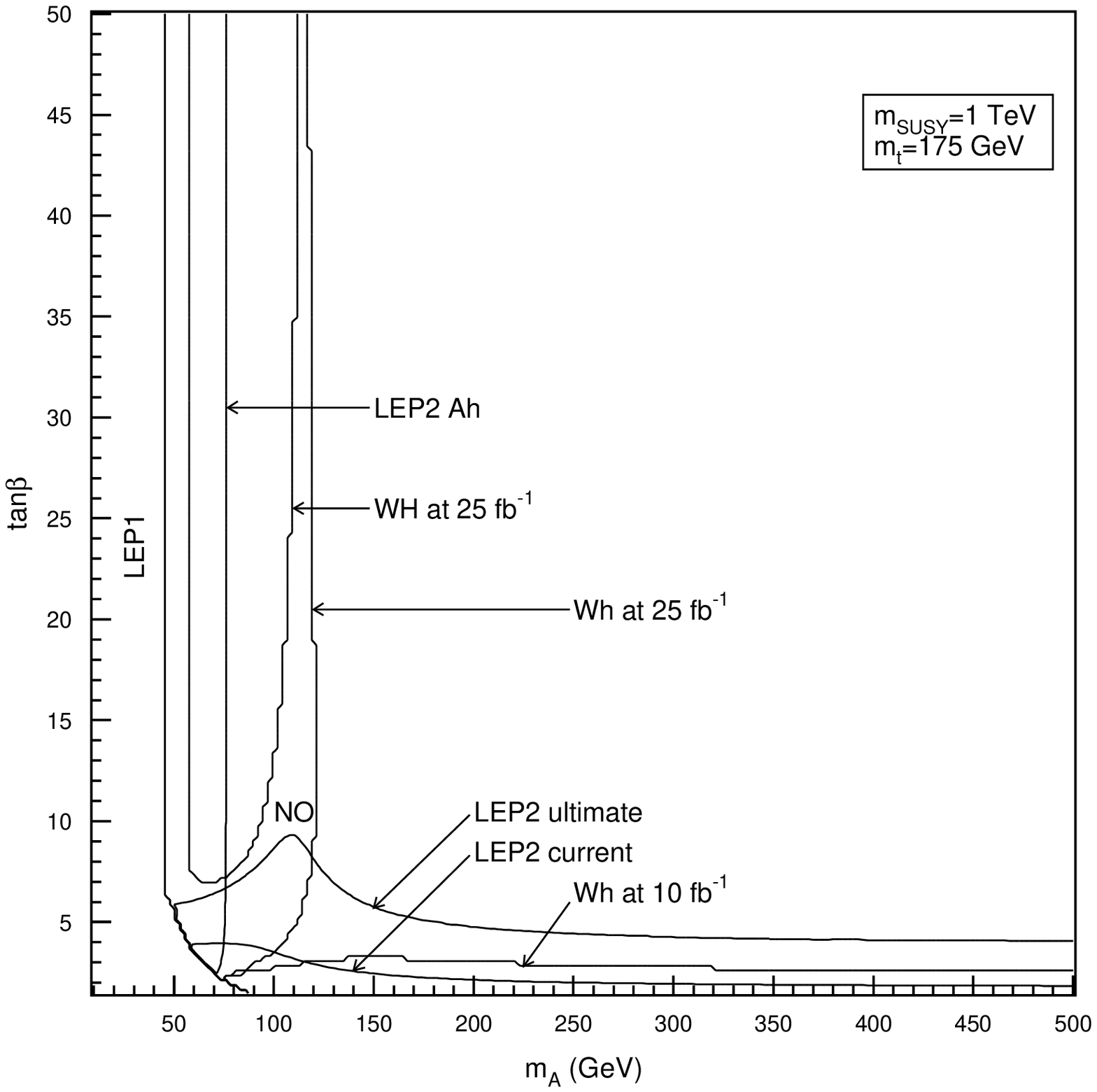}}
\caption{}
\label{FIG1}
\end{figure}

Our first results are presented in Fig.\ \ref{FIG1} for the MSSM $m_A\
vs.\ \tan\beta$ plane, where all SUSY mass parameters have been set to 1
TeV except the $A$-parameters which have been set to zero, 
and $m_{top}=175$ GeV. The region labelled LEP1 is excluded because
the rate for $Z\to Z^*h,\ hA$ and $H^+H^-$ would violate limits placed
on Higgs masses from LEP1 runs at the $Z$-pole. The region below the
contour labelled ``$LEP2$ current'' yields a $Zh$ or $ZH$ signal in
excess of the $ZH_{SM}$ rate for $m_{H_{SM}}=88$ GeV.  In the region
below the ``$LEP2$ ultimate'' contour, the signal ought to be detectable
by LEP2 experiments sensitive to $m_{H_{SM}}=107$ GeV at $\sqrt{s}=200$
GeV.  The region to the left of the contour labelled ``$LEP2\ Ah$'' 
is excluded by current LEP2
experiments searching for $Z^*\to Ah$.  The ultimate LEP2 $Ah$ reach is
only a slight extension of the current $LEP2\ Ah$ contour and is not
shown.  The reach via $H^+H^-$ production lies well within the region
probed by $Ah$ searches. Finally, there may exist a region of this
parameter plane at high $\tan\beta$ that is excluded by $b\bar{b}h$,
$b\bar{b}H$ and $b\bar{b}A$ searches at the Tevatron, where the Higgs
bosons decay to $\tau\bar{\tau}$; refined calculations of this
are in progress\cite{drees}.

The reach of the Fermilab Tevatron experiments for $\ell\nu b\bar{b}$ 
events with 10 fb$^{-1}$ of integrated luminosity corresponds to the
region below the contour marked ``$Wh$ at 10 fb$^{-1}$''. If the 
integrated luminosity is increased to 25 fb$^{-1}$, then the region
to the right of the contour labelled ``$Wh$ at 25 fb$^{-1}$'' can be
seen. In addition, at low values of $m_A$, the region interior to the
contour labelled ``$WH$ at 25 fb$^{-1}$'' can be seen. 

We find a window of non-observability (labelled NO)
at $m_A\simeq 115$ GeV where {\it none} of the SUSY Higgs bosons can be
seen at LEP2 or the Fermilab Tevatron collider. 
In this region, both $h$ and $H$ may contain a significant portion of the
SM Higgs scalar, so that $\sin(\alpha+\beta)$ 
is significantly smaller than unity and $\sigma(Wh)$ is suppressed, 
while $H$ is somewhat 
heavier than 120~GeV~\cite{footnote}.
This window is similar to but smaller than the one noted in
Refs. \cite{baer,gunion,zwirner,barger,atlas,cms} that also occurs at
the CERN LHC.  Most or all of the window may be filled in at the CERN
LHC by high luminosity running, and by searches for $\tau\bar{\tau}$,
$\mu\bar{\mu}$ or $4b$ modes from Higgs production.  If the Tevatron 
integrated luminosity is increased to 30 fb$^{-1}$, the $WH$ region
increases towards lower $\tan\beta$, but most of the unobservable domain
persists even at this higher integrated luminosity, and is not filled in
even by our projection of the ultimate reach of LEP2.  The NO window
persists even for an integrated luminosity of 50 fb$^{-1}$.  A similar
plot for $\mu =-1000$ GeV was constructed, but was indistinguishable
from Fig.\ \ref{FIG1}. Finally, to test the robustness of our
conclusions, we re-mapped our contours assuming an increase of 3 GeV in
all the Higgs masses throughout parameter space. The resulting plot was
qualitatively similar to the one presented in Fig.\ \ref{FIG1}, with a
modest increase in the size of the window of non-obervability.

\subsection{Minimal Supergravity Model (mSUGRA)}

\begin{figure}
\centerline{\epsfbox{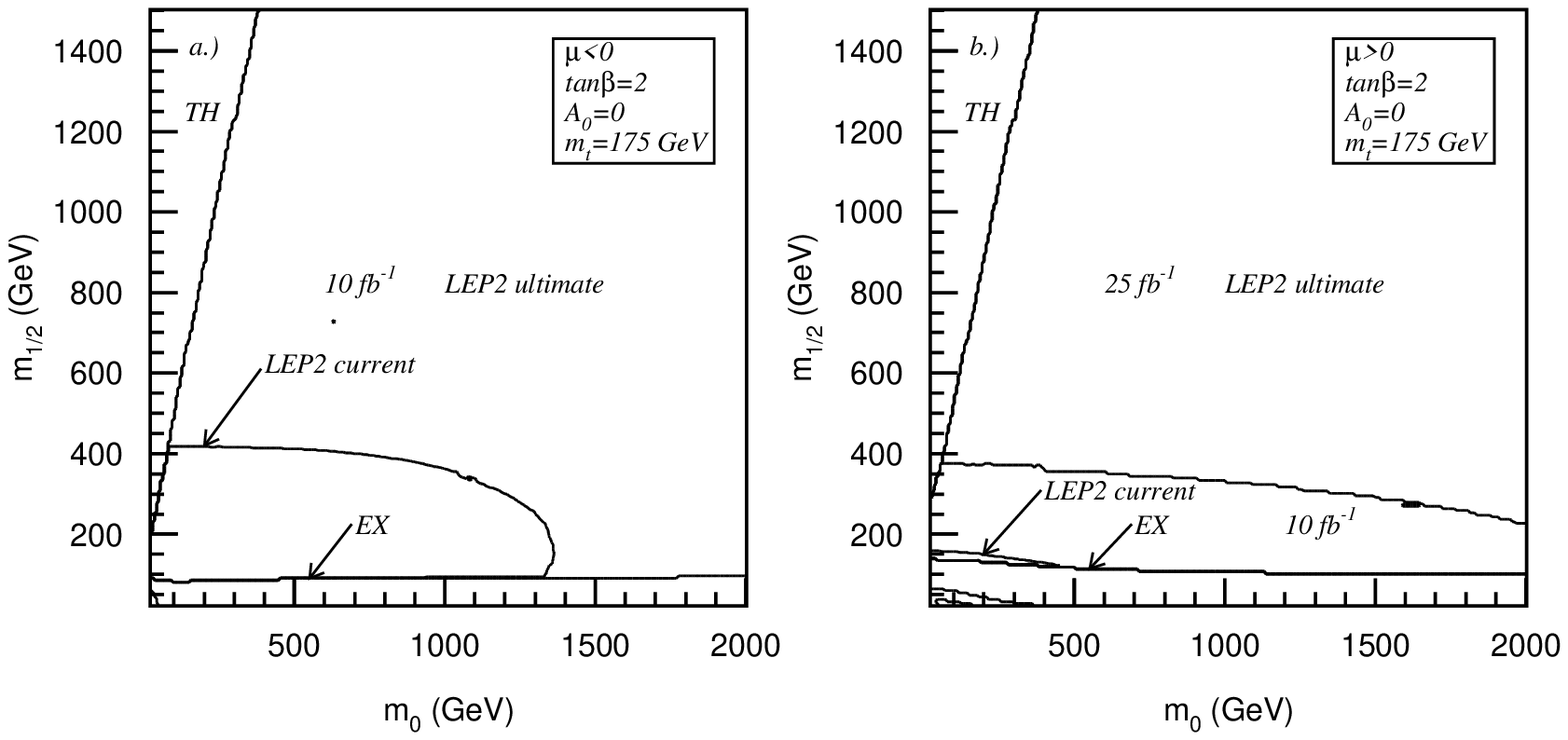}}
\vspace {-3.5in}
\caption{}
\label{FIG2}
\end{figure}

A similar analysis of SUSY Higgs observability can be made within the 
context of the mSUGRA model, except that in this case there exists the 
possibility of light superpartners which can affect both the Higgs boson
masses and branching fractions.  The results are presented in 
Fig.\ \ref{FIG2} for $\tan\beta =2$ and for {\it a}) $\mu <0$ and
{\it b}) $\mu >0$. The TH region is excluded either due to a lack of 
appropriate electroweak symmetry breaking, or because the lightest
neutralino is not the lightest SUSY particle (LSP).
The EX region is excluded by the negative results from LEP2  searches
for chargino pair 
production, which require $m_{\tw_1} > 85$ GeV\cite{lep2wino}. 

The Higgs boson search at  LEP2 presently excludes the region below the
contours labelled ``$LEP2$ current''. We see that for $\tan\beta =2$ and
$\mu <0$, the current LEP2 Higgs bound rules out a significant portion
of mSUGRA parameter space up to $m_{1/2}\simeq 400$ GeV, corresponding
to $m_{\tg}\simeq 1000$ GeV! However, for $\mu >0$ it  
gives only a slight additional excluded area beyond the chargino bound.
For both parameter planes for $\tan\beta =2$, experiments at LEP2 will
ultimately probe the entire parameter space shown, so that these will
either  discover a Higgs boson or rule out mSUGRA if $\tan\beta$ happens 
to be small\cite{bbmt}.

The Fermilab Tevatron collider has no reach for Higgs bosons with 
2 fb$^{-1}$ of integrated luminosity, but ought to be able to explore 
the entire plane of Fig.\ \ref{FIG2}{\it a} for $\mu <0$ 
with 10 fb$^{-1}$. For $\mu >0$, the Tevatron 
reach is to $m_{1/2}\simeq 300-400$ GeV for 10 fb$^{-1}$, while the entire
plane can be explored with 25 fb$^{-1}$ of integrated luminosity
that might be accumulated at the TeV33 upgrade of the Tevatron. 
The entire $\tan\beta =2$ parameter space can be 
explored by LEP2 and a 25 fb$^{-1}$ Tevatron even if we allow
$m_h\to m_h+3$ GeV everywhere.

\begin{figure}
\centerline{\epsfbox{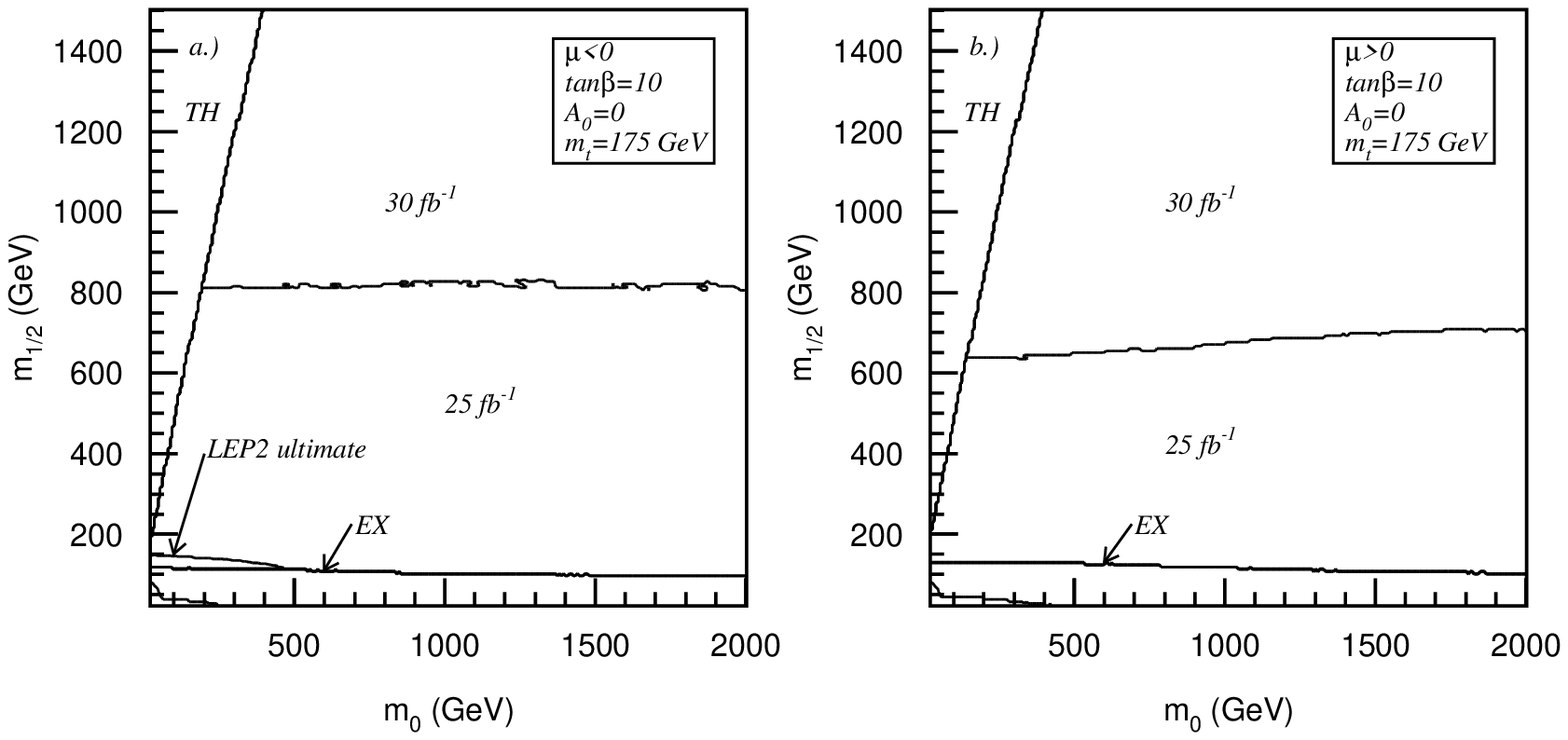}}
\vspace {-3.5in}
\caption{}
\label{FIG3}
\end{figure}

An intermediate value of $\tan\beta =10$ is shown for the mSUGRA model
in Fig.\ \ref{FIG3}{\it a} and \ref{FIG3}{\it b} for $\mu <0$ and $\mu >0$,
respectively. As $\tan\beta$ increases, $m_h$ generally
increases\cite{ltanbl}, and in fact the current LEP2 SM Higgs search
imposes no additional constraints beyond those from other SUSY particle
searches.  In fact, the LEP2 ultimate reach lies only slightly beyond
the current chargino mass exclusion contour, making Higgs searches at
LEP2 difficult for larger values of $\tan\beta$.

At $\tan\beta =10$, the Fermilab Tevatron experiments 
have {\it no reach} for Higgs bosons
with 10 fb$^{-1}$ of integrated luminosity. However, if an
integrated luminosity of 25 fb$^{-1}$ can be accumulated, then the Tevatron
Higgs search may probe $m_{1/2}$ values as high as $m_{1/2}\simeq 800$
GeV (650 GeV) for $\mu <0$ ($\mu >0$) via the $Wh$ channel. 
(These values change to $m_{1/2}\sim 300-500$ GeV if $m_h\to m_h+3$ GeV). 
An increase
in integrated luminosity to 30 fb$^{-1}$ allows complete coverage of the
parameter space shown.

\begin{figure}
\centerline{\epsfbox{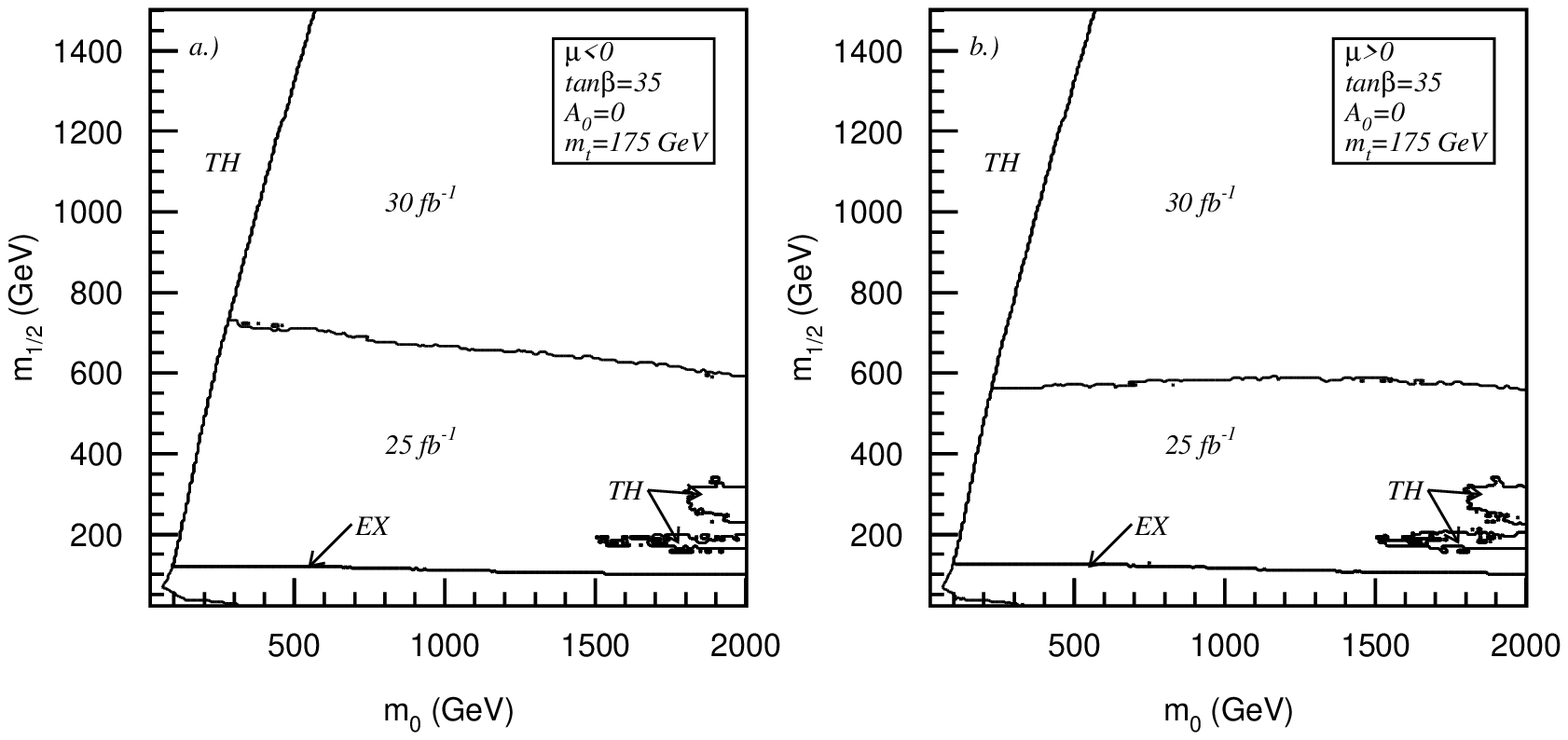}}
\vspace {-3.5in}
\caption{}
\label{FIG4}
\end{figure}

Finally, we show in Fig.\ \ref{FIG4}{\it a} and {\it b} the corresponding 
mSUGRA reach plots for $\tan\beta =35$. In this case, there are some 
additional TH excluded regions at large $m_0$ and $m_{1/2}\sim 100-300$ 
due to lack of radiative EWSB, though the exact computation of this region 
is numerically somewhat sensitive. 
In this large $\tan\beta$ case, there will be {\it no reach}
by LEP2 Higgs searches beyond the regions already excluded by 
current chargino searches. However, Fermilab Tevatron experiments can explore
via $Wh$ searches up to $m_{1/2}\simeq 600-700$ GeV with 
25 fb$^{-1}$ of integrated luminosity ($m_{1/2}\simeq 300-400$ GeV if 
$m_h\to m_h+3$ GeV), and they can probe once again the entire parameter 
space shown with 30 fb$^{-1}$ of integrated luminosity.  
From these results, we conclude that the Fermilab Tevatron collider with
30 fb$^{-1}$ of integrated luminosity ought to either discover a light 
Higgs boson, or can exclude the mSUGRA model. We do not expect our 
results to change significantly with changes in the $A_0$ parameter.

\subsection{Gauge Mediated SUSY Breaking Model (GMSB)}

\begin{figure}
\centerline{\epsfbox{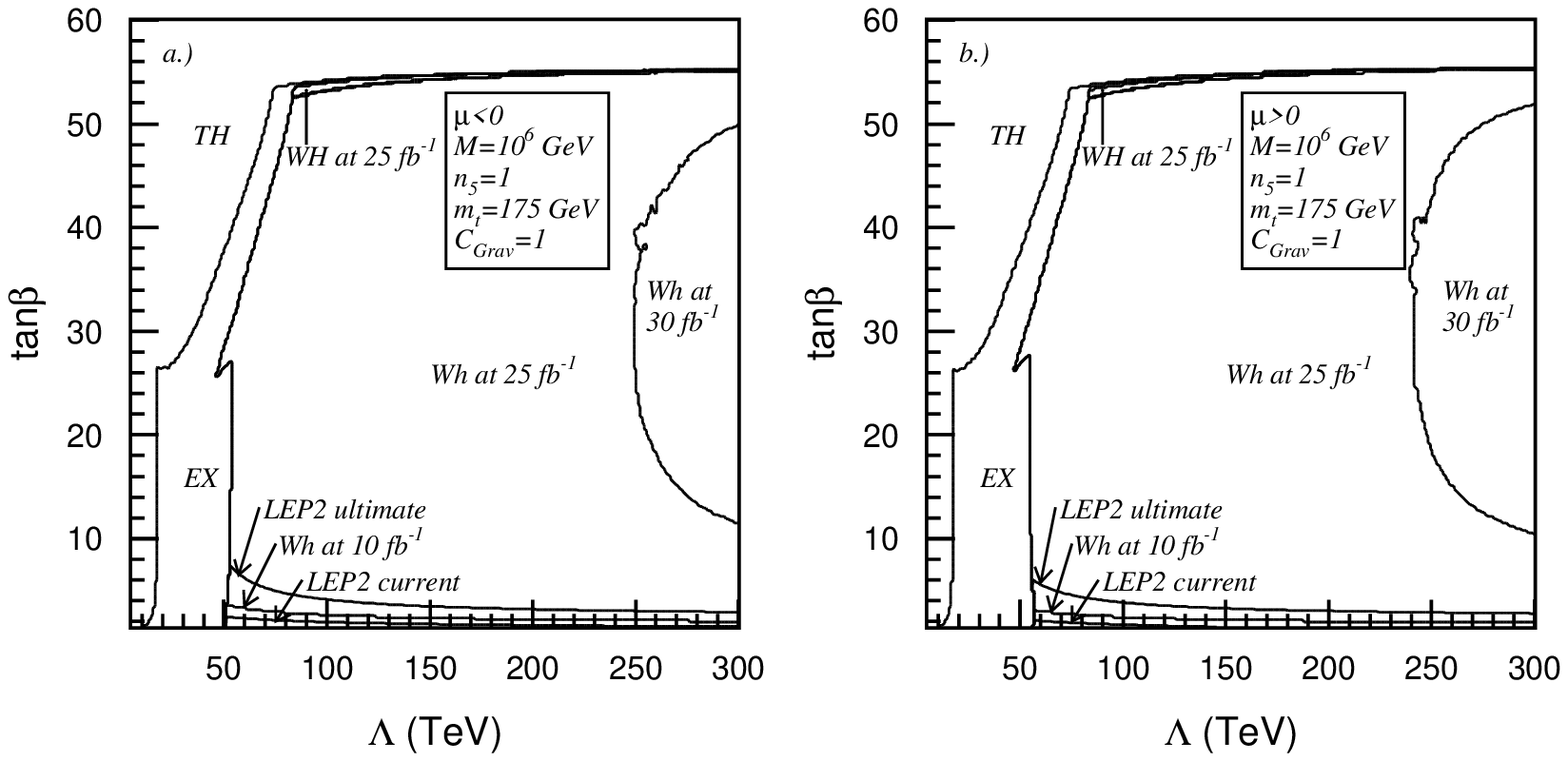}}
\vspace {-3.5in}
\caption{}
\label{FIG5}
\end{figure}

Results for the GMSB model are shown in Fig.\ \ref{FIG5}, in the
$\Lambda\ vs.\ \tan\beta$ plane for {\it a}) $\mu <0$ and {\it b}) $\mu
>0$.  We take the messenger scale $M_{mes}=1000$ TeV and first assume
the messenger sector is comprised of just a single set $n_5=1$ of
``quark'' and ``lepton'' superfields in a $5+\bar{5}$ representation of
$SU(5)$. The TH regions are excluded because they do not lead to 
proper radiative EWSB:
in these regions, either $m_A^2<0$ or $m_{\ttau_R}^2<0$. The EX regions
are excluded by the ALEPH bound $m_{\tz_1}>71$ GeV  on the unstable
neutralino which decays via $\tz_1\rightarrow \gamma\tilde{G}$ from
searches
for $e^+e^-\to\gamma\gamma\eslt$~\cite{alephmz1}, or by
$m_{\ttau_1}>60$ GeV, from searches for staus in GMSB
models\cite{alephstau}. The D0 and CDF experiments at 
the Tevatron~\cite{DZERO},
from a non-observation of di-photon events with large $\eslt$ have
inferred a lower bound of about 150~GeV on the chargino mass: this
excludes essentially the same region as the ALEPH neutralino bound.

The current bound from LEP2 due to Higgs boson searches is shown as the
region below the contour labelled ``LEP2 current'' in the very low $\tan\beta$
region of each plot. The LEP2 ultimate reach is for the area below the contour
labelled ``$LEP2$ ultimate'', and can only probe parameter space with
$\tan\beta <4-7$. 

Experiments at the Fermilab Tevatron will explore the region below the
``$Wh$ at 10 fb$^{-1}$'' contour with 
$10$ fb$^{-1}$ of integrated luminosity.
By increasing the integrated luminosity to $25$ fb$^{-1}$, 
almost the entire parameter space should be explorable via Higgs boson
searches. The exception is the region at $\Lambda\agt 250$ TeV
and $\tan\beta\sim 10-50$. This additional region, however, can be covered with
an integrated luminosity of $30$ fb$^{-1}$. Finally, there is
a narrow sliver of region at $\tan\beta\simeq 52-54$ which is not seeable
at the Tevatron via $Wh$ searches; however, most of this sliver 
is accessible via $WH$ production at $25$ fb$^{-1}$. 

We have also again examined the effect of setting $m_h\to m_h+3$ GeV
for the GMSB model.
The most obvious impact is on the size of the region that
cannot be explored with 25~fb$^{-1}$ of data: the boundary of this
region shifts from $\Lambda \sim 250$~TeV to $\Lambda \sim 150$~TeV for
both signs of $\mu$, and furthermore, this region then extends down to
$\tan\beta \sim 6$.

\begin{figure}
\centerline{\epsfbox{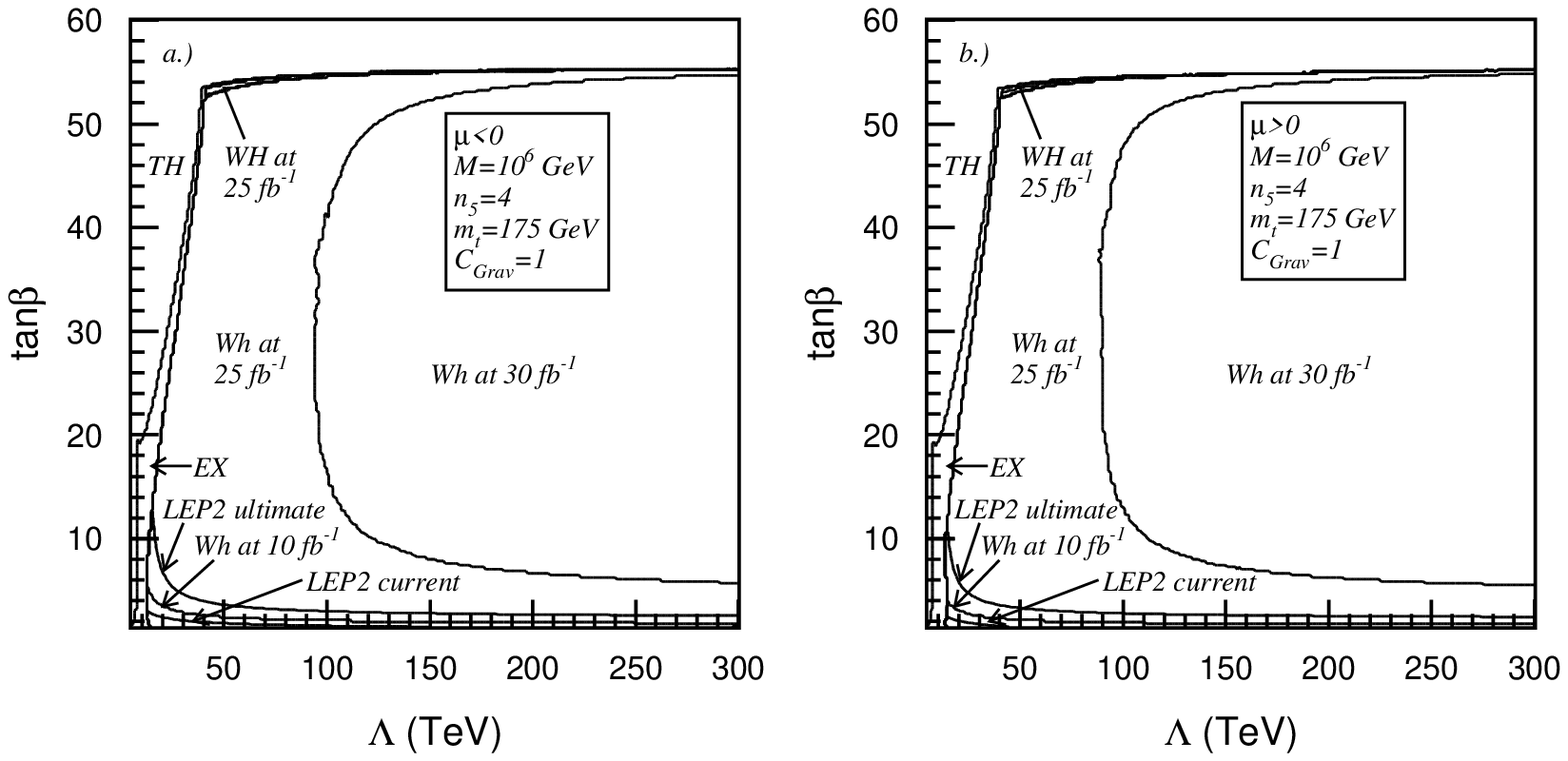}}
\vspace {-3.5in}
\caption{}
\label{FIG6}
\end{figure}

Finally, we have also explored the Higgs boson reach for the case
$n_5=4$, the largest value of $n_5$ for which couplings remain
perturbative up to a large scale when $M_{mes}$ is 1000~TeV.  The EX
region may be somewhat underestimated~\cite{pedro} since we obtain it using the
same bound $m_{\tz_1}> 71$~GeV used in the $n_5=1$ case, in conjunction
with the bound on $m_{\ttau_1}$.  In this case,
$m_h$ is typically larger for a given $\Lambda$ and $\tan\beta$ value
than for the $n_5 =1$ case. These results are shown in Fig.\ \ref{FIG6}.
The overall structure of the plot is similar to the results from Fig.\
\ref{FIG5}, except that the $Wh$ reach for 25 fb$^{-1}$ is considerably
smaller.  However, the whole parameter plane shown can once again be
covered by 30 fb$^{-1}$ of integrated luminosity.

\section{Conclusions}

The lightest neutral scalar Higgs boson is lighter than $120-125$~GeV in a
wide variety of interesting models commonly used for phenomenological
studies of weak scale supersymmetry. This observation, together with the
fact that the most careful study to date yields a SM Higgs boson reach
of about 120~GeV at proposed luminosity upgrades of the Tevatron,
motivated us to translate this reach to the corresponding reach in three
classes of SUSY models: the MSSM with $m_{SUSY}=1$~TeV, the mSUGRA
model, and finally, the GMSB model. While there are viable models where
the $\sim 125$~GeV limit on the lightest of the neutral Higgs scalars can be
significantly surpassed, these models generally contain additional fields (most
commonly Higgs singlets) which serve no specific purpose, and hence are
less attractive theoretically.

Assuming that the conclusion about the reach for the SM Higgs boson
holds up to closer scrutiny~\cite{workshop}, we find that it should be
possible to discover at least one of the Higgs bosons over the entire
parameter space of the mSUGRA and GMSB models, and over most of the
parameter space of the MSSM, provided Tevatron experiments can
accumulate an integrated luminosity of 25-30~fb$^{-1}$.  The large
integrated luminosity is crucial for an exploration of the parameter
space: if the integrated luminosity is ``just'' 10~fb$^{-1}$ the signal
is detectable over a much smaller range of parameters.  Specifically,
for the case of the MSSM, experiments at TeV33 should be able to probe
the entire model parameter space given 25 fb$^{-1}$ of integrated
luminosity, except for a window of unobservability around $m_A\sim 115$
GeV. For the case of the mSUGRA model, both LEP2 and TeV33 should be
able to probe the entire parameter space for low $\tan\beta$; for higher
$\tan\beta$, the reach of LEP2 will be limited, and Tevatron experiments
will need 30 fb$^{-1}$ of integrated luminosity to probe the entire
model parameter space. For GMSB models, the ultimate reach of LEP2 is
limited to low values of $\tan\beta$, while the Fermilab Tevatron will
need $\sim 30$ fb$^{-1}$ of integrated luminosity to probe the entire
parameter space. This conclusion is insensitive to the specific value of
$n_5$ as long as the messenger scale $M_{mes}$ is not too large.

In summary, if future luminosity upgrades of the Tevatron are able to
accumulate an integrated luminosity of 25-30~fb$^{-1}$, experiments at
these facilities should be able to probe most of the parameter space of
these models via a search for $p\bar{p} \to Wh (H) \to \ell\nu b\bar{b}$
events. While a discovery of a signal would be very exciting, it would
not, by itself, allow us to conclude that we had discovered SUSY. If
instead Tevatron experiments do accumulate the required data sample and
conclusively find no signal for the Higgs boson, it would exclude most
of the parameter space of currently popular SUSY models, and cause us to
rethink our ideas of how weak scale supersymmetry might be realized in
nature.

%
\acknowledgments
We thank M. H. Reno for collaboration in the early stages of this work.
This work was supported in part by the U.~S. Department of Energy
under contract numbers DE-FG-FG03-94ER40833 and DE-FG02-97ER41022.
%

\begin{center}
{\bf NOTE ADDED IN PROOF}
\end{center}
After submission of this paper for publication, another paper\cite{carena}
appeared in which the detectability of SUSY Higgs bosons was 
examined within the MSSM. In Ref. \cite{carena}, it was pointed out 
that the region of the $m_A-\tan\beta$ plane (our Fig.\ 2) where there is an
observable Higgs boson signal shows some dependence on the sign of
$A_t\mu$. 
This does not show up in our Figure 2 since we have set $A_t=0$. 
We have checked and confirm (for $A_t=1$~TeV) that the signal shows 
this sign dependence: the reach region is somewhat smaller for
$\mu < 0$ than for $\mu >0$. For $m_A \geq 200$~GeV and large $\tan\beta$
($\geq 40-45$), the signal goes from being just above observability (for
$\mu >0$) to just below observability (for $\mu < 0$) with 25~$fb^{-1}$.
With 30~$fb^{-1}$, the signal dependence on the sign of $A_t\mu$ 
still makes some difference, but only for a limited range of 
$m_A$ around 120-150~GeV.
%

\newpage
%

\centerline{\bf \large{Figure Captions}}
\begin{description}

\item[Fig.\ 1]
Number of events expected after cuts at the Fermilab Tevatron 
for 10 fb$^{-1}$ of integrated luminosity for the $WH_{SM}$ signal
(solid) and SM background (dashes), as a function of $m_{H_{SM}}$.
The crosses denote points from the calculations of Kim {\it et al.},
Ref. \cite{kuhlmann}. The curves denote our fit to these points.
\item[Fig.\ 2]
Regions of the MSSM $m_A\ vs.\ \tan\beta$ plane accessible to current and 
future Higgs boson searches at LEP2 and the Fermilab Tevatron. We take 
$m_{top}=175$ GeV and set $m_{SUSY}=1$~TeV.
\item[Fig.\ 3]
Regions of mSUGRA model parameter space accessible to current and future
Higgs boson searches at LEP2 and the Fermilab Tevatron. We take 
$m_{top}=175$ GeV, $A_0=0$ and $\tan\beta =2$.
\item[Fig.\ 4]
Regions of mSUGRA model parameter space accessible to current and future
Higgs boson searches at LEP2 and the Fermilab Tevatron. We take 
$m_{top}=175$ GeV, $A_0=0$ and $\tan\beta =10$.
\item[Fig.\ 5]
Regions of mSUGRA model parameter space accessible to current and future
Higgs boson searches at LEP2 and the Fermilab Tevatron. We take 
$m_{top}=175$ GeV, $A_0=0$ and $\tan\beta =35$.
\item[Fig.\ 6]
Regions of GMSB model parameter space accessible to current and future
Higgs boson searches at LEP2 and the Fermilab Tevatron. We take 
$m_{top}=175$ GeV, $M_{mes}=1000$~TeV and $n_5=1$.
\item[Fig.\ 7]
Regions of GMSB model parameter space accessible to current and future
Higgs boson searches at LEP2 and the Fermilab Tevatron. We take 
$m_{top}=175$ GeV, $M_{mes}=1000$~TeV and $n_5=4$.

\end{description}


\newpage

\newpage

\newpage

\newpage

\newpage

\newpage

\newpage


\begin{references}
\bibitem{footnote} In the window of non-observability, $h$ and $H$ tend
to be close in mass. If they are close enough so that $m_{jj}$ falls
into a common bin, the signal from the two could be combined, and may
then satisfy our observability criterion. We have not attempted to do so
here, but an analysis which uses variable bins may optimize the signal.
\end{references}
\end{document}